# Unanticipated proximity behavior in ferromagnet-superconductor heterostructures with controlled magnetic noncollinearity

L. Y. Zhu[1], Yaohua Liu[1], F. S. Bergeret[2,3,4], J. E. Pearson[1], S. G. E. te Velthuis[1], S. D. Bader[1,5] and J. S. Jiang[1,*]

[1]Materials Science Division, Argonne National Laboratory, Argonne, IL 60439, USA

[2]Centro de Física de Materiales (CFM-MPC), Centro Mixto CSIC-UPV/EHU, E-20018 San Sebastián, Spain

[3]Donostia International Physics Center (DIPC), Manuel de Lardizabal 4, E-20018 San Sebastián, Spain

[4]Institut für Physik, Carl von Ossietzky Universität, D-26111 Oldenburg, Germany

[5]Center for Nanoscale Materials, Argonne National Laboratory, Argonne, IL 60439, USA

**Abstract**

Magnetization noncollinearity in ferromagnet-superconductor (F/S) heterostructures is expected to enhance the superconducting transition temperature ($T_c$) according to the domain-wall superconductivity theory, or to suppress $T_c$ when spin-triplet Cooper pairs are explicitly considered. We study the proximity effect in F/S structures where the F layer is a Sm-Co/Py exchange-spring bilayer and the S layer is Nb. The exchange-spring contains a single, controllable and quantifiable domain wall in the Py layer. We observe an enhancement of superconductivity that is nonmonotonic as the Py domain wall is increasingly twisted via rotating a magnetic field, different from theoretical predictions. We have excluded magnetic fields and vortex motion as the source of the nonmonotonic behavior. This unanticipated proximity behavior suggests that new physics is yet to be captured in the theoretical treatments of F/S systems containing noncollinear magnetization.

Proximity enables multicomponent composites to embrace antagonistic properties whose mutual influence gives rise to a wealth of intriguing phenomena. For example, singlet superconductivity and ferromagnetism are mutually exclusive in homogeneous bulk materials, but they can coexist at the interface of ferromagnet-superconductor (F/S) heterostructures [1]. Singlet Cooper pairs penetrate only a few nanometers into the F layer due to the strong exchange field, leading to short-range proximity effects, such as an oscillatory critical temperature ($T_c$) in S/F superlattices [2, 3], and $\pi$ state S/F/S Josephson junctions [4, 5]. Long-range proximity effects could also arise provided there is magnetization noncollinearity at the F/S interface, where the spin rotation by the inhomogeneous exchange field converts singlet Cooper pairs into triplets [6]. A unique signature predicted for triplet superconductivity is the suppression of $T_c$ due to the leakage of the long-range triplet pairs into F [7]. However, it has also been shown that, in the case of F/S interface with a Néel-wall-like noncollinearity, although the long-range triplets are present, they have no influence on $T_c$; superconductivity is enhanced due to a reduction of the effective exchange field experienced in the domain wall region by the singlet pairs [8].

Prior experimental observations of the superconducting spin switch effect [9, 10] and domain wall superconductivity [11] are qualitatively consistent with theoretical predictions of F/S proximity effects involving nonuniform ferromagnets [12-14]. Definitive comparison between theories and experiments, however, is problematic. The magnetic domain structures in experimental samples can be rather complex, and most experiments assume or infer the magnetic configurations. The localized enhancement of superconductivity near domain walls necessarily means that not only the existence, but also the specific arrangements, of domain walls influence the proximity effects in F/S systems [15]. Another complication is the magnetostatic stray fields that invariably accompany inhomogeneous magnetization distributions such as domain walls and sample edges. The stray fields could suppress superconductivity by the classical orbital effect or by dissipative vortex motion, or could enhance conductance by vortex pinning. It has been argued that some experimental observations of spin switch and inverse spin switch effects could be alternatively explained via domain-state dominated mechanisms [16, 17]. In order to have a better understanding of F/S proximity effects in the presence of inhomogeneous magnetization, it is imperative to design experiments with samples possessing a well-defined and properly characterized magnetic structure [18, 19].

In this Letter, we report the experimental observation of a *nonmonotonic enhancement* of superconductivity with the increase of magnetic noncollinearity in a F/S system containing a single, controllable, and quantifiable noncollinear magnetic structure. Our results cannot be accounted for with the singlet domain-wall superconductivity theory [8] that predicts a monotonic enhancement of superconductivity with increasing magnetic noncollinearity, nor with the triplet superconducting spin switch theory [7] that predicts a suppression of superconductivity due to the long-range triplet ordering in the presence of magnetic noncollinearity. This unanticipated proximity effect suggests that there may be new physics yet to be captured in the present theories of F/S proximity effect with noncollinear magnetization.

We used Nb for the S layer, and an exchange-spring (ES) Sm-Co/Py bilayer as the F layer that provides noncollinear magnetization. In an ES bilayer, due to the interfacial exchange coupling between the magnetically hard (Sm-Co) and soft (Py) layers, a spiral spin structure can be achieved in the soft layer with negligibly small anisotropy when a magnetic field is applied at an angle from the anisotropy axis of the hard layer [see Fig. 1(a)]. The spin spiral is similar to an in-plane Bloch domain wall; its pitch (noncollinearity) and handedness (chirality) are governed by the applied field and its directional history [20, 21]. Our F/S samples have the configuration

MgO/Cr(20 nm)/Sm-Co(50 nm)/Py($t$)/Nb(30 nm)/Cr(2 nm), with $t$=10 or 33 nm. The Sm-Co layer was epitaxially grown on Cr-buffered MgO (110) single-crystal substrates to ensure a single uniaxial anisotropy axis along the MgO [001] direction. Our samples differ from the polycrystalline SmFe/Py/Nb structures of Ref. [22] in that the well-defined Sm-Co anisotropy enables us to quantify the magnetization noncollinearity and to adopt measurement conditions that definitively rule out possible experimental artifacts. We used deposition setup and conditions as reported in our previous work [20] to prepare the epitaxial Sm-Co layers, and the Py, Nb, and Cr layers were deposited subsequently at room temperature. It is worth noting that the 2-nm-thin Cr capping layer is nonmagnetic and therefore is not affecting the adjacent Nb layer magnetically. The epitaxial growth of Sm-Co was verified using x-ray diffraction, and the magnetic and superconducting properties were characterized utilizing magnetometry and electrical transport measurements, respectively.

Figure 1(b) shows the normalized resistance $R(\theta)/R(0°)$ of the $t$=10 nm F/S sample as a function of the angle $\theta$ (0° to 360°, then back to 0°) between the Sm-Co saturation magnetization and magnetic field $H$ directions. The measurements were performed at 4.5 K, where the sample resistance is ~50% of the normal state resistance, during sample rotation in a series of magnetic fields. The measurements utilized a 4-probe geometry with an excitation current of 10 μA applied in-plane and perpendicular to the Sm-Co easy axis. The $R(\theta)$ curves are symmetric with respect to $\theta$=180°. The sample has the highest $R$ at $\theta$=0° when the magnetization is collinear. $R$ initially decreases, by as much as 35%, at $\theta$ values (~106°) that are quite insensitive to the magnitude of the rotating field. Further rotating the field toward $\theta$=180° increases $R$. It is worth noting that $R(\theta=0°)$ is always higher than $R(\theta=180°)$. The results are strikingly different from those obtained at 10 K, above the superconducting transition, as shown in Fig. 1(c). At 10 K, $R(\theta)$ is also symmetric with respect to $\theta=180°$, but $R$ initially increases with $\theta$, reaching a maximum before decreasing as $\theta$ increases toward 180°. The $R$ variation at 10 K is less than 0.2% and the $\theta$ values at which $R$ reaches maximum is field dependent. While $R(\theta)$ at 10 K is dominated by the anisotropic magnetoresistance (AMR) in the Py layer, at 4.5 K it is due to current shunting as the Nb layer enters the superconducting transition. We note that the initial decreases in $R(\theta)$ at $T_c$ resemble those reported in Ref. [22] for SmFe/Py/Nb structures, however, the AMR behaviors at 10 K are opposite due to the different choices of the measurement current direction. Therefore, $R(\theta)$ measured at $T_c$ is not associated with the AMR. The variation in $R$ at a fixed temperature in the resistive transition suggests a nonmonotonic enhancement of superconductivity in Nb as the applied field is rotated away from alignment with the Sm-Co easy axis. Based on the slope of the $R(T)$ curve at the midpoint of the resistive transition, we estimate that a 35% decrease in $R$ would correspond to an increase of $T_c$ by ~10 mK.

To verify that the resistance change at 4.5 K during field rotation is indeed related to changes in superconductivity, we examined the angle dependence of the superconducting critical current $I_c(\theta)$. We carried out transport measurements in the Corbino geometry using lithographically patterned 250 nm-thick Nb electrodes placed at the center of the F/S samples, ~1 mm from the edges (see Fig. 2 inset). When the thick Nb electrodes become superconducting, the measurement current is confined within the ring-shaped region between the two Nb electrodes, and the transport measurements are not affected by stray fields from the sample edges. Shown in Fig. 2 is the $I_c(\theta)$ of the $t=10$ nm F/S sample measured at 4.5 K using the Corbino geometry. $I_c$ was extracted from the $I$-$V$ curves using the criterion that a voltage exceeding 1 µV signifies the superconducting-normal transition. From the saturated state where all spins are collinear ($\theta=0°$), $I_c$ first increases as the spin spiral winds up, reaching a maximum at $\theta \sim 102°$ before decreasing as the field is rotated toward $\theta=180°$. This nonmonotonic enhancement of $I_c$ is a further indication that increasing magnetic noncollinearity affects superconductivity.

However, a proximity effect due to the noncollinear spin spiral in the ES during field rotation is but one possible mechanism for the nonmonotonic resistance change in the superconducting transition region and the $I_c$ enhancement. Changes in any source of magnetic field acting on the S layer could also lead to similar behaviors due to altered orbital pair breaking or vortex pinning. To ascertain that the observed nonmonotonic enhancement of superconductivity during field rotation is a proximity effect in nature, we compared the field-angle dependence of the superconducting transitions in pairs of F/S and ferromagnet-insulator-superconductor F/I/S structures having identical F and S layers. The F/I/S structures have a 15-nm thick intrinsic Si layer that becomes insulating at low temperatures. Without transmission of Cooper pairs between the F and S layers, the F/I/S structures cannot exhibit any proximity effect. Shown in Fig. 3(a) is the normalized $R(\theta)/R(0°)$ curve of the $t=33$ nm F/S sample measured at 4.4 K when a 0.15 T in-plane field was rotated in the sample plane from $\theta=0°$ to $360°$ and back to $0°$. Between $0°$ and $\sim 150°$, the up- and down-sweep branches of the $R(\theta)$ curve are nonhysteretic and show gradual fall and rise similar to those seen in the $t=10$ nm F/S sample. At $\theta>150°$, the $R(\theta)$ curve is hysteretic and the up- and down-sweep branches each has a sharp peak around $\theta=180°$. On the other hand, the $R(\theta)$ curve of the corresponding $t=33$ nm F/I/S sample [Fig. 3(b)] remains constant at most $\theta$ values, except for the two sharp peaks at locations similar to those of the F/S sample. $T_c$ of the F/I/S sample is $\sim 2$ K higher because of the absence of a proximity effect. The flat background in $R(\theta)$ of the F/I/S sample rules out the possibilities that $H$ is misaligned with the sample plane, or that the S layer is in the vortex flow regime at all. This result is also confirmed with experiments where two identical F/S samples, having the excitation current applied in orthogonal directions, were measured together but no difference was found. The two sets of sharp peaks in the $R(\theta)$ curves are the signatures of out-of-plane stray fields due to multidomain formation. As the spin spiral winds up in a rotating magnetic field, the magnetic exchange energy increases, and the spin spiral reverses its chirality to reduce the stored magnetic exchange energy when the field rotation exceeds a critical angle [20]. During the reversal, lateral magnetic domains of opposite chiralities coexist in the Py layer, producing out-of-plane stray fields at the domain walls that suppress superconductivity in the S layer. The presence of the two resistance peaks at similar locations is evidence that the F/S and F/I/S samples have similar magnetic domain structures during chirality reversal. Since the process of chirality switching is irreversible, the presence of the domain wall stray fields is necessarily indicated by an irreversibility in $R(\theta)$. Therefore, the absence of hysteresis in $R(\theta)$ of the F/S sample between 0° and 150° demonstrates that the sample remains singledomain within that angular range, and that the observed nonmonotonic behavior of $R(\theta)$ could not have been the result of domain wall stray fields. Taken together, the experimentally observed nonmonotonic resistance change in the resistive transition and $I_c$ enhancement in the ES-based F/S samples represent a nontrivial, proximity effect where superconductivity in S is modified by the magnetic noncollinearity in F via the transmission of Cooper pairs.

The large AMR effect in the Py layer allows us to quantitatively determine the spin profiles and correlate their evolution with the superconducting properties in the same samples and under the same field history. We modeled the $R(\theta)$ curves in Fig. 1(c) with micromagnetic calculations in which the Sm-Co/Py bilayer is treated as a series of exchange-coupled sublayer slices; and the sublayer slices contribute to the total resistance as resistors in-parallel [20, 23].

The best-fit $R(\theta)$ curves are shown as solid lines in Fig. 1(c). A single set of micromagnetic parameters reproduce all the features in the $R(\theta)$ curves measured at all field values, and the micromagnetic parameters from the best-fit ($K_{Py}$=1.13×10$^5$ ergs/cm$^3$, $A_{Py}$=1.45×10$^{-6}$ ergs/cm, $M_{Py}$=743 emu/cm$^3$ and $K_{Sm-Co}$=5.0×10$^7$ ergs/cm$^3$, $A_{Sm-Co}$=1.2×10$^{-6}$ ergs/cm, $M_{Sm-Co}$=442 emu/cm$^3$) are also close to the respective nominal bulk values. The fit also yields the spin depth profile $\varphi(x)$ for any direction and magnitude of the external field. The robustness of the fit gives us confidence about the uniqueness of the $\varphi(x)$. In Fig. 4(a) inset we plot $\varphi(x)$ for a 0.8 T in-plane field applied at a series of angles $\theta$. The angular range over which the spin spiral spans monotonically increases with $\theta$. However, the bulk of the spin spiral resides close to the Sm-Co/Py interface. Shown in Fig. 4(a) is the calculated spin noncollinearlity $\Delta\varphi$ across the top 1 nm of Py near the Py/Nb interface as a function of $\theta$ for a series of applied field values. The magnitude of $\Delta\varphi$ is rather small, reaching ~2.4° at $\theta$=180° for a 1 T field. Although for low fields $\Delta\varphi$ varies nonmonotonically with $\theta$, the $\theta$ values at which $\Delta\varphi$ reaches maximum are 128° for $H$=0.3 T, and 168° for $H$=0.6 T.

With the quantitatively determined spin profiles, which are unchanged when the Nb layer enters the superconducting state due to the much smaller superconducting gap energy compared with the ferromagnetic exchange energy, we can compare the experimentally observed non-monotonic superconductivity enhancement with expectations from proximity effect theories. The singlet-based Néel domain wall superconductivity theory gives the $T_c$ enhancement $\Delta T/T_c \sim (\Delta\varphi)^2$, where $\Delta\varphi$ is the total rotation angle of the exchange field within the superconducting coherence length $\xi_s$ [8]. A naive extension of the theory would give $\Delta\varphi$ as the total rotation angle of the exchange field within the coherence length $\xi_F = \sqrt{4\hbar D_F/h}$ in the F layer ($D_F$ and $h$ are the diffusion constant and exchange energy of the F layer, respectively). $\xi_F$ is ~1 nm for Py. Taking into account the calculated $\Delta\varphi$ shown in Fig. 4(a), we find that a singlet-based proximity effect would give a nonmonotonic superconductivity enhancement at low applied fields. However, while $\Delta\varphi$ generally increases with increasing field values and reaches maximum values at rotation angles that are field dependent, the experimentally observed superconductivity enhancement peaks at $\theta \sim 106°$ and the magnitude of the enhancement decreases with increasing field. The experimental results in our Sm-Co/Py/Nb structures do not appear to be attributable to the singlet-related domain wall proximity effect.

We also considered explicitly the generation of triplet components of the superconducting condensate by incorporating the quantitatively determined magnetization non-collinearity. We solve the Usadel equation for the quasiclassical Green's function (GF) in a S/F structure [6, 7]. The F layer has the magnetization profile shown in the inset of Fig. 4(a) and the S/F interface is described by the Kupryianov-Lukichev boundary conditions [24]. To calculate $I_c$ flowing in the S layer, a phase gradient in $y$ direction is considered

$$\check{f}_S = \hat{f}_S(\hat{\tau}_1 \cos ky + \hat{\tau}_2 \sin ky) \quad , \tag{1}$$

where $\hat{f}_S$ is the anomalous GF matrix in spin space and $\hat{\tau}_i$ are Pauli matrices in particle-hole space. To simplify the problem, we assume (i) the thickness of the S layer $d_S$ is smaller than $\xi_s$, so the GF in the S layer can be integrated over thickness, and (ii) temperatures close to $T_c$ to linearize the equations. In order to obtain $I_c$, a self-consistent problem needs to be solved [25, 26]. From the linearized Usadel equation and the self-consistent equation, the critical temperature $T_c$ can be calculated for a given value of the phase-gradient parameter $k$. For temperatures close to $T_c$, the temperature dependence of the order parameter $\Delta(T)$ can be approximated as $\Delta(T)^2 \sim 2\pi^2(T_c^2 - T^2)$ [27]. Finally, the current density is given by
9

$$j(\alpha) = \frac{\pi}{4} e v_0 D_S kT Tr \sum_{\omega_n > 0} \hat{f}_S^2(\alpha) \quad , \qquad (2)$$

where $v_0$ is the normal density of states at the Fermi level, $D_S$ is the diffusion coefficient of the S-layer, $\alpha = D_S k^2/2$ and $\omega_n$ are the Matsubara frequencies. $I_c$ is determined by the maximum value of $j$. Using the spin profiles shown in Fig. 4 inset for the $t=10$ nm F/S structure, we have calculated the expected $I_c(\theta)$ behaviors at various $T$, with a wide range of values for the parameters $\gamma_b$, $\xi_s$, and $r$, where $r = \rho_S/\rho_F$ is the resistivity ratio of the S and F layers, and $\gamma_b = R_b S/\rho_F \xi_F$ is the Kupryianov-Lukichev interface parameter ($R_b$ is the S/F interface resistance and $S$ is the interface area). Several characteristic $I_c(\theta)$ curves are shown in Fig. 4(b) ($T_c(\theta)/T_c(0)$ curves show similar trend). The general trend of an initial decrease in the calculated $I_c$ with increasing $\theta$, while similar to the $T_c$ suppression predicted for the superconducting triplet spin valve in Ref. [7], is opposite to the experimentally observed nonmonotonic $I_c$ enhancement shown in Fig. 2. We presently cannot reconcile the differences between the experimental observations and theoretical calculations. We note, however, a similar nonmonotonic enhancement of superconductivity was previously reported for SmFe/Py/Nb structures [22], although the polycrystalline nature of those samples prevented an unambiguous conclusion. We emphasize that, by using epitaxial ES to create well-defined magnetic configurations and having definitively ruled out spurious field sources with the F/I/S structure, we have established a firm set of experimental observations of an unanticipated proximity behavior that is nontrivial. These experimental observations are the constraints that need to be accounted for when developing theoretical treatments of F/S interfaces containing magnetic noncollinearity.

In conclusion, we quantitatively examined the superconducting proximity effect in epitaxial ES F/S heterostructures, with Sm-Co/Py as the ferromagnet and Nb as the superconductor. We find that the enhancement of superconductivity shows a non-monotonic dependence on the noncollinearity of the magnetization structure. We have demonstrated that proximity effect is the underlying mechanism of the experimental observations. The measured dependence of the superconducting transition on magnetization noncollinearity cannot be explained by either the singlet domain wall superconductivity or the triplet theories. We hope the observation of this unconventional effect will stimulate further refinement of F/S proximity effect theories.

Work at Argonne and use of the Center for Nanoscale Materials are supported by the U.S. Department of Energy, Office of Science, Office of Basic Energy Sciences, under Contract No. DE-AC02-06CH11357. The work of F.S.B was supported by the Spanish Ministry of Economy and Competitiveness under Project No. FIS2011-28851-C02-02.

* jiang@anl.gov

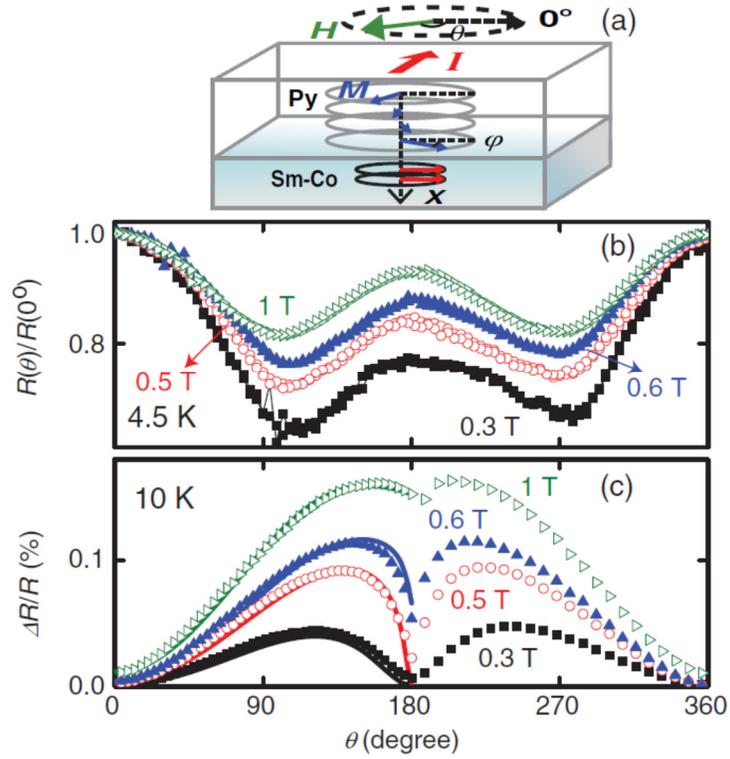

FIG. 1 (a) Schematic spin spiral diagram of an ES with the definitions of $\theta$, $\varphi$, $x$, and the measurement geometry. Normalized resistance of the $t$=10 nm F/S sample as a function of $\theta$ with various magnetic fields at (b) 4.5 and (c) 10 K (symbols). The solid lines are fits to the data and $\Delta R/R=[R(\theta,H)-R(0,0)]/R(0,0)$.

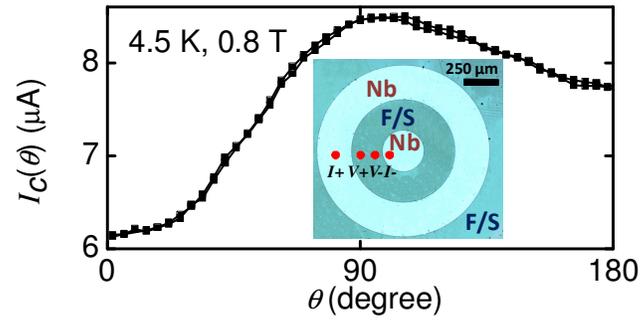

FIG. 2 $I_c(\theta)$ of the $t$=10 nm F/S sample measured at 4.5 K with a 0.8 T in-plane field in the Corbino geometry as shown in the inset micrograph.

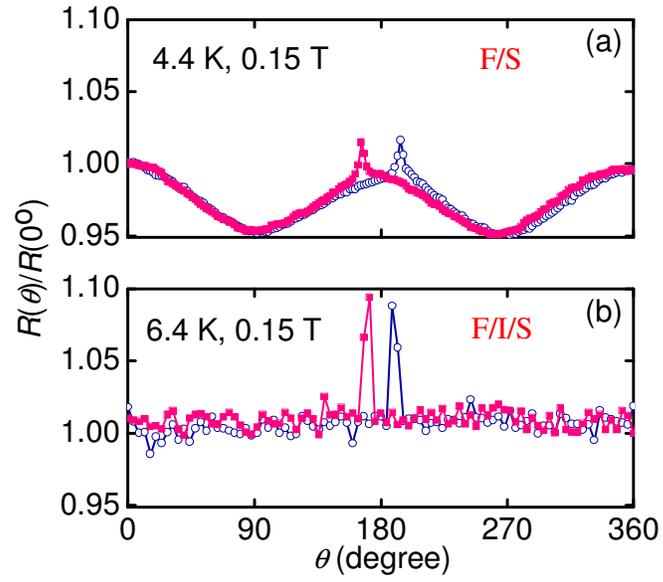

FIG. 3 Normalized $R(\theta)/R(0^o)$ loops measured under a 0.15 T in-plane rotating field for the $t$=33 nm (a) F/S sample at 4.4 K. (b) F/I/S sample at 6.4 K. $\theta$ changes from $0^o$ to $360^o$ (open circles) and back to $0^o$ (solid squares).

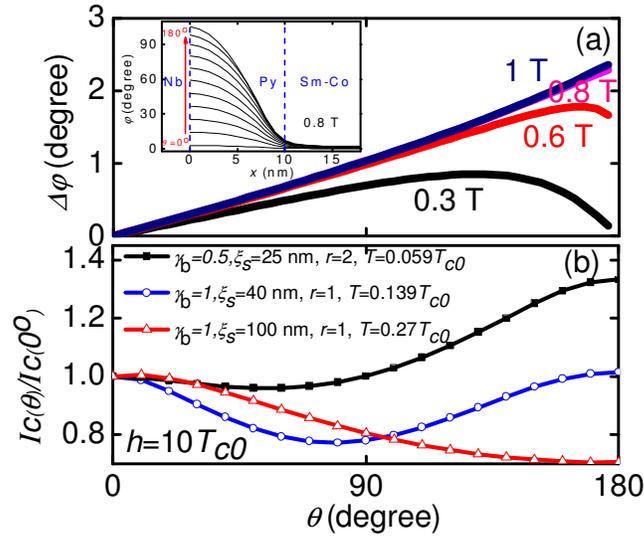

FIG. 4 (a) Spin noncollinearity $\Delta\varphi=\varphi(x=0\text{ nm})-\varphi(x=1\text{ nm})$ in the $t=10$ nm F/S sample at various in-plane fields. Inset: spin rotation angle $\varphi(x)$ for a 0.8 T in-plane $H$ applied at various directions from 0° to 180°. (b) Normalized $I_c(\theta)$ for the $t=10$ nm F/S sample when rotating a 0.8 T in-plane field. $T_{c0}$ is the intrinsic critical temperature of Nb without the F layer. For all curves $T/T_c \sim 0.83$.